\documentclass[preprint,12pt,eqsecnum]{aastex}

\newcommand{\Cs}{{\cal C}}
\newcommand{\drho}{\delta \rho}
\newcommand{\dv}{\delta v}
\newcommand{\dep}{\delta p}
\newcommand{\deta}{\delta \eta}
\newcommand{\dzeta}{\delta \zeta}
\newcommand{\dL}{\Delta L}
\newcommand{\Li}{(L^{-1})}
\newcommand{\dR}{\Delta R}
\newcommand{\om}{\omega}
\newcommand{\si}{\sigma}
\newcommand{\pr}{\partial_r}
\newcommand{\pz}{\partial_z}
\newcommand{\ppr}{\partial r}
\newcommand{\ppz}{\partial z}
\newcommand{\pf}{\partial_\phi}

\newcommand{\LL}{{\cal L}}
\newcommand{\zme}{\xi}
\newcommand{\oor}{r^{-1}}
\newcommand{\LLv}{\LL_{\rm v}}
\newcommand{\dV}{\delta V}
\newcommand{\dVz}{\delta V_0}
\newcommand{\dVo}{\delta V_1}
 
\newcommand{\occ}{orthonormal cylindrical coordinates}
\newcommand{\linea}{------}

\slugcomment{Accepted for publication by The Astrophysical Journal}
\shorttitle{Perturbations of Viscous Rotating Fluids}

\begin{document}
\title{On The Perturbations Of Viscous Rotating Newtonian Fluids}

\author{Manuel Ortega-Rodr\'{\i}guez }
\affil{Department of Applied Physics and Gravity Probe B, \\
Stanford University, Stanford, CA 94305--4090}
\email{manuel@leland.stanford.edu}
\and
\author{Robert V. Wagoner}
\affil{Department of Physics, Stanford University, Stanford, CA 94305--4060}   
\email{wagoner@leland.stanford.edu}                                                              

\begin{abstract}
The perturbations of weakly-viscous, barotropic, non-self-gravitating, Newtonian rotating fluids are analyzed via a single partial differential equation. The results are then used to find an expression for the viscosity-induced normal-mode complex eigenfrequency shift, with respect to the case of adiabatic perturbations. However, the effects of viscosity are assumed to have been incorporated in the unperturbed (equilibrium) model. This paper is an extension of the normal-mode formalism developed by Ipser \& Lindblom for adiabatic pulsations of purely-rotating perfect fluids. The formulas derived are readily applicable to the perturbations of thin and thick accretion disks. We provide explicit expressions for thin disks, employing results from previous relativistic analyses of adiabatic normal modes of oscillation. In this case, we find that viscosity causes the fundamental p- and g-modes to grow while the fundamental c-mode could have either sign of the damping rate.
\end{abstract}  

\keywords{hydrodynamics --- accretion, accretion disks --- stars: oscillations}


\section{INTRODUCTION}

The problem of perturbations of rotating fluids is of considerable interest in astrophysics. These perturbations can give us information not only about stars and accretion disks 
(e.g., \markcite{FI}Friedman \& Ipser 1992; \markcite{NW}Nowak \& Wagoner 1992; \markcite{KFM}Kato et al.~1998), but also about the properties of the compact object at the center of the latter (e.g., \markcite {PSWL}Perez et al.~1997).

Traditionally, this problem has been approached in two different ways.
The classical way has been to use a Lagrangian displacement formalism (\markcite{LO}Lynden-Bell \& Ostriker 1967). An alternative method uses two scalar potentials (proportional to the perturbations of the pressure and gravitational potential) within the framework of Eulerian fluctuations (\markcite{P}Poincar\'e 1885; \markcite{IM}Ipser \& Managan 1985; \markcite{M}Managan 1985). In the spirit of the latter approach, Ipser \& Lindblom (\markcite{ILb}1991b, hereafter IL; \markcite{IL2}1992) have developed an elegant formalism to analyze adiabatic pulsations of perfect Newtonian (as well as relativistic) fluids that are stationary, axisymmetric, and purely rotating (i.e., have no meridional component of velocity).
They look for normal-mode solutions and reduce the problem to two coupled second-order partial differential equations for the two potentials.

The inclusion of viscosity in such a system is an important step in the modeling of various types of astrophysical phenomena (e.~g., \markcite{ILa}Ipser \& Lindblom 1991a). 
The purpose of this paper is to explore some of the consequences arising from the presence of weak viscosity, taking into consideration that its effect is two-fold: 
a) it changes the value of the unperturbed fluid variables that characterize the equilibrium configuration, and b) it changes the form of the differential equations that govern the evolution of the perturbations. However, we assume that step (a) has already been carried out. We will work within the Cowling approximation (i.e., neglecting gravitational perturbations), which reduces the system to a single differential equation. This approximation is often appropriate (\markcite{C}Cowling 1941; \markcite{IL2}Ipser \& Lindblom 1992), and is certainly valid for most accretion disks. 

In section 2 we present a derivation of the governing partial differential equation.
In section 3 we use this result to obtain a formula for the viscosity-induced shift of the complex eigenfrequencies of the perturbations.
This enables us, in section 4, to compare our results with those of Ipser \& Lindblom (\markcite{ILa}1991a), who perform the calculation for the simpler case of pure uniform rotation of the equilibrium model (which then can contain no effects of viscosity). 
In section 5, we apply our formalism to thin accretion disks. Explicit integral expressions (involving the eigenfunctions) are obtained for the contributions to the growth (or damping) rate of a mode. We then evaluate the order of magnitude of that rate for the fundamental modes. 
Except where indicated, we shall employ the notation and conventions of IL.


\section{GENERALIZED IPSER--LINDBLOM FORMALISM}

The equations of motion of a Newtonian fluid are 
\begin{eqnarray}
  \partial_{t} \rho + \nabla_{\! a} ( \rho v^{a} ) & = & 0 \; ,  \\
  \rho (\partial_t v^a + v^b \nabla_{\! b} v^a) & = &
   - \nabla^a p + \rho \nabla^a \Phi + \nabla_{\! b} \sigma^{a b} \; .\label{cm}
\end{eqnarray}
In the cases we shall consider, $\Phi$ represents a specified fixed, axisymmetric gravitational potential (which does not necessarily obey the Newtonian field equation). The viscous stress tensor $\si^{ab}$ is defined by 
\begin{equation}
  \sigma^{a b} \equiv \eta (\nabla^a v^b + \nabla^b v^a) +
     g^{ab}  (\zeta - \case{2}{3} \eta) \nabla_{\! c} v^c \; .
\end{equation}
In this initial investigation, we adopt a generalized barotropic equation of state which specifies the pressure $p$ as well as the shear and bulk viscosity coefficients $\eta$ and $\zeta$ as functions of the mass density $\rho$ only. In these equations, $\partial_t$ and $\nabla_{\! a}$ represent the partial derivative with respect to time and the spatial Euclidean covariant derivative. Below, $\partial_a = {}_{,a} = \partial/\partial x^a$ indicate spatial partial derivatives, with $x^a = r,\phi,z$. The Euclidean metric $g_{ab}$ and its inverse $g^{ab}$ are used to raise and lower tensor indices.

We assume that the viscosity coefficients are small enough that the effects of the viscous term in equation ({\ref{cm}}) can be treated using standard perturbation techniques. To be explicit, we assume that the term $ \nabla_{\! b} \sigma^{a b} $ has a relative magnitude of order $\alpha \ll 1$ compared to the dominant terms in the equation. 
This implies that the viscosity coefficients have a magnitude of order $\alpha\rho_*L_*^2 \Omega_*$, where $\rho_*$, $L_*$, and $\Omega_*$ refer to the typical density, length, and frequency scales of the system. (For an accretion disk, $L_*$ corresponds to the thickness of the disk.) In this way we can treat $\alpha$ as a constant perturbation parameter with respect to the inviscid ($\alpha = 0$) case, and work to first order in $\alpha.$

The presence of viscosity will cause a change in the equilibrium values of the dynamical variables. In particular, it can induce meridional components ($r,z$) of the velocity field.
Using the noncoordinate basis of \occ \ (with unit vectors $r^a, \phi^a$, and $z^a$; and
$g_{ij} = \delta_{ij}$), the latter can be expressed as
\begin{equation}
v^a = r \Omega \phi^a + v^r r^a + v^z z^a \; .
\end{equation}
The last two terms on the right-hand-side of this equation are assumed to have a relative magnitude of order $\alpha$ with respect to the first term. (For instance, this is true for a thin accretion disk.)

The equations for the evolution of the Eulerian perturbations (in the Cowling approximation) are
\begin{equation}
  \partial_{t} \drho + v^a \nabla_{\! a} \drho + \drho \nabla_{\! a} v^a +    
      \nabla_{\! a} ( \rho \dv^{a}) = 0  \label{e1} 
\end{equation}
and
\begin{eqnarray}
  \partial_t \dv^a + v^b \nabla_{\! b} \dv^a + \dv^b \nabla_{\! b} v^a 
  & = & -{\nabla^a \dep \over \rho} + {\drho \nabla^a p \over \rho^2}
       - {\drho \nabla_{\! b} \sigma^{a b} \over \rho^2} \nonumber \\
& & \mbox{}+ \rho^{-1}\Big{\{} \nabla_{\! b} \big{[} \eta (\nabla^a \dv^b + \nabla^b \dv^a)]
       + \nabla_{\! b}[\deta (\nabla^a v^b + \nabla^b v^a)] \nonumber \\ 
& & \mbox{}+ \nabla^a [(\zeta - \case{2}{3} \eta) \nabla_{\! c} \dv^c] + \nabla^a [(\dzeta - \case{2}{3} \deta) \nabla_{\! c}v^c]{\Big{\}}} \; .  \label{e2}
\end{eqnarray}
Specifying the functions
\begin{equation}
  \Gamma (r,z) \equiv {\rho \over p}{dp \over d\rho}\; , \label{dep}
\end{equation}
\begin{equation}
  \mu (r,z) \equiv {\rho \over \eta}{d\eta \over d\rho} \label{deta}
\end{equation}
effectively closes the system of differential equations.
(It turns out that the contributions of $\dzeta$ are of higher order in $\alpha$.)
The inviscid case has been studied in an elegant fashion by IL in more generality. (They included gravitational perturbations and allowed a general equation of state, but neglected meridional velocities.)
The purpose of this paper is to explore the effects of viscosity on their results under the restrictions of the Cowling approximation and our generalized barotropic equations of state.

We look for normal mode solutions to the equations, which (because of the stationarity and axisymmetry of the equilibrium configuration) have a time dependence of the form $e^{i \si t}$ and an azimuthal angular dependence of the form $e^{i m \phi}$. The constant $\si$ is the inertial frequency of the mode and the axial mode number $m$ is an integer. Let us express equation (\ref{e2}) in a way that shows which part depends explicitly on $\alpha$ and which does not:
\begin{equation}
  (L^a_{\ b} + \alpha \, \dL^a_{\ b}) \dv^b  = (R^a + \alpha \, \dR^a) \dep \; . \label{lr} \end{equation} 
The derivation of the $\alpha = 0$ terms, $R^a$ and $L^a_{\ b}$, is found in IL. 
Here we merely quote their results. 
The operator $R^a$ is equal to $- \nabla^a \rho^{-1}$, where throughout this paper the operator $\nabla^a$ is to affect all the terms appearing to its right.
The operator $L^a_{\ b}$ is related to the operator $Q^a_{\ b}$ of IL by  $L^a_{\ b} = i (Q^{-1})^a_{\ b}$, where (in our \occ $\ r, \phi, z$)
\begin{equation}
Q = {1 \over \om(\om^2 - \kappa^2)}\left[\matrix{\om^2 & -2i\Omega \om & r (\Omega^2)_{,z} \cr
   {i \om \kappa^2 / 2 \Omega} & \om^2 & i r \om \Omega_{,z} \cr
   0  & 0  & \om^2 - \kappa^2 \cr}\right] \; . \label{q}
\end{equation}
The corotating frequency is $\om = \si + m\,\Omega (r,z)$, and $\kappa^2 = 2\Omega(2\Omega + r \Omega_{,r})$ is the square of the radial epicyclic frequency. We note that $\om(\om^2-\kappa^2)$ is the determinant of $Q^{-1}$. (Also note that IL use a cylindrical {\it{coordinate}} basis and that their roles of $\si$ and $\om$ are switched with respect to ours.) The derivation of the expressions for $\alpha\dL^a_{\ b}$ and $\alpha\dR^a$ is straightforward and can be found in the Appendix.

The key to the procedure is the fact that $Q$ is purely algebraic. This allows us to solve for the velocity perturbation:
\begin{equation}
\dv^c = \Li^c_{\ a}(R^a + \alpha\,\dR^a)\dep - \alpha\,\Li^c_{\ a} \dL^a_{\ b} \dv^b \; .
\end{equation}
Thus, to the same order in $\alpha$,
\begin{equation}
   \dv^c \approx \Li^c_{\ a} (R^a + \alpha \, \dR^a) \dep
     - \alpha \, \Li^c_{\ a} \dL^a_{\ b} \Li^b_{\ d} R^d \dep \; . \label{deltav}
\end{equation}
Terms of order $\alpha^2$ and higher have been dropped (and will be ignored from now on).

Taking into account equation (\ref{dep}), we can express equation (\ref{e1}) in the following way:
\begin{equation}
(\om - i\alpha\Cs)(\rho /p\Gamma)\dep - i\nabla_{\! a}(\rho\dv^{a}) = 0 \; , \label{cont}
\end{equation}
where the operator
\begin{equation}
\alpha\Cs \equiv r^{-1}(r v^r)_{,r}+ v^z_{,z} + v^r \pr + v^z \pz \; .
\end{equation}
Throughout this paper, the operators $\pr$ ($=\partial/\partial r$) and $\pz$ ($=\partial/\partial z$) are to affect all the terms appearing to their right.
Substituting equation (\ref{deltav}) in equation (\ref{cont}), we obtain
\begin{equation}   
\nabla_{\! a}(\rho Q^a_{\ b}\nabla^b\rho^{-1}\dep) + {\om \rho \over p \Gamma}\dep
= \alpha {\Big{[}} \nabla_{\! a}(\rho Q^a_{\ b} \dR^b \dep) - i \nabla_{\! a}(\rho Q^a_{\ b} \dL^b_{\ c}Q^c_{\ d}\nabla^d \rho^{-1}\dep)+i\Cs{\rho\over p\Gamma}\dep{\Big{]}}\; . \label{mas}
\end{equation} 
We have thus reduced the original problem to a single partial differential equation for one unknown, $\dep$. 
Once solved, $\drho$, $\deta$, and $\dv^a$ can be obtained via equations (\ref{dep}), (\ref{deta}), and (\ref{deltav}), respectively.
Our master equation (\ref{mas}) plays the same role as equation (22) of IL, which has the same structure within the Cowling approximation. 


\section{EIGENFREQUENCY SHIFTS}

Following IL, for reasons that will become apparent below we will use a new variable:
\begin{equation}
   \dV \equiv \dep/\rho\om \; .
\end{equation}
With this definition we can express equation (\ref{mas}) as
\begin{equation}
   (\LL + \alpha\,\LLv) \dV = 0 \; , \label{master2}
\end{equation}
where
\begin{eqnarray}
\LL & \equiv & \nabla_{\! a} \rho Q^a_{\ b} \nabla^b \om
    + \om^2\rho^2/p\Gamma \; , \label{Ls} \\
\LLv & \equiv & -\nabla_{\! a} \rho Q^a_{\ b} \dR^b \om \rho + i \nabla_{\! a} \rho Q^a_{\ b} \dL^b_{\ c} Q^c_{\ d} \nabla^d\om -i\Cs\om\rho^2/p\Gamma \; . 
\end{eqnarray}
The operators $\LL$ and $\LLv$ depend upon $\alpha$ implicitly through the ($r$ and $z$ dependent) properties $v^a, \rho, p, \eta, \zeta$ of the unperturbed model as well as through the eigenfrequency $\sigma$. However, we assume that the equilibrium model already incorporates the effects of viscosity, at least through first order in $\alpha$. Thus we shall be concerned only with the effects of viscosity on the perturbations, which then changes the operator    
$\LL$ only through its dependence on $\sigma$. 

Expanding the eigenfrequency $\sigma = \sigma_0 + \Delta\sigma + \ldots$ and the eigenfunction $\dV = \dVz + \alpha\,\dVo + \ldots$, we can express equation (\ref{master2}) in the following way, to first order in $\alpha$:
\begin{equation}
  \left[\LL + (\partial\LL/\partial\sigma)\Delta\sigma + \alpha\LLv\right]_{\sigma_0}
(\dVz + \alpha\dVo) = 0 \; .
\end{equation}
All operators are now evaluated at the adiabatic eigenfrequency $\sigma_0$.
Using the fact that $\LL\dVz = 0$, we then obtain 
\begin{equation} 
  \alpha\,\LL\dVo + \Delta\sigma(\partial\LL/\partial\sigma)\dVz
  + \alpha\,\LLv\dVz  = 0 \; . \label{anterior}
\end{equation}

It turns out that with the boundary condition $p=0$ on a smooth, compact, finite area surface of the body, $\LL$ is Hermitean (IL). This implies that
\begin{equation}
  \int\dVz^*\,\LL\,\dVo\, d^3x = \int(\LL\,\dVz)^*\,\dVo\, d\,^3x = 0 \; .
\end{equation}
Thus if we multiply equation (\ref{anterior}) by  $\dVz^*$ and integrate, we obtain 
\begin{equation}     
  \Delta\sigma\langle\partial\LL/\partial\sigma\rangle
  + \alpha\,\langle\LLv\rangle = 0 \; ;
\end{equation}     
where
\begin{equation}
   \langle {\cal O} \rangle \equiv \int \dVz^* \, {\cal O} \, \dVz \, d\,^3x \; , 
\end{equation}    
only for operators {$\cal O$} evaluated at $\sigma=\sigma_0$. 

Then we find (from the structure of $\LL$ and $\LLv$) that  
\begin{equation}
{\rm{Re}}(\Delta\si) = 0 \: , \qquad i{\rm{Im}}(\Delta\si) = -\frac{\alpha\langle\LLv\rangle}{\langle\partial\LL/\partial\si\rangle} \; . \label{finali}
\end{equation} 
Equation (\ref{finali}) allows us to compute the viscosity-induced shift of the eigenfrequencies, provided we are given an equilibrium model.
Equation (\ref{anterior}) can then be used to obtain the eigenfunction correction $\dVo$.


\section{PERTURBATIONS OF UNIFORMLY ROTATING BODIES}

In order to compare our results with those obtained by \markcite{ILa}Ipser \& Lindblom (1991a), we need to invoke their assumptions of uniform rotation and vanishing of meridional velocity (i.e., we need to set $\Omega_{,r} = \Omega_{,z} = v^r = v^z = 0$). Thus, there are no effects of viscosity on their equilibrium model since the viscous shear tensor vanishes. 
In addition, since we do not consider self-gravitational effects, we need to set $\delta \Phi = 0$ in their formalism.
Under these assumptions, equation (\ref{finali}) reproduces their results [equations (27), (28), and (29)] for the imaginary part of the frequency induced by viscous effects. 
[Their $\tau^{-1}$ corresponds to our Im($\Delta \si$), and their canonical energy $E$, to our quantity $(\om/2)\langle\partial\LL/\partial\si\rangle$.]
When dealing with differentially rotating bodies, our approach is more straightforward than an extension of theirs, which employs a Lagrangian approach.
  

\section{APPLICATION TO ACCRETION DISKS}

The results obtained so far can be applied to the special case of a thin accretion disk. 
Its physical properties allow us to retain relatively few terms from the ones contained in equation (\ref{finali}). Adiabatic oscillations of such accretion disks have been studied within 
a radial WKB approximation (\markcite{NW}Nowak \& Wagoner 1992; \markcite{PSWL}Perez et al.~1997; \markcite{SW}Silbergleit \& Wagoner 1999) $\lambda_r \ll r_*$, with $\lambda_r\gtrsim h_*$ for most of the low-lying modes. The length scales $2h_*$ and $r_*$ are the typical thickness and radial size of the disk region of interest, and $\lambda_r$ is the radial scale of variation of the eigenfunction. In addition, we shall only consider values of the axial mode number $|m|\lesssim 1$.

The system is characterized by the small parameter $\epsilon\equiv h_*/r_*$.  In order to use our equations we will also need a model for the unperturbed disk, which we take to be similar to the one developed by Kita \& Klu\'zniak (\markcite{KK}1998). In agreement with their results, we take $v^r \sim \alpha\epsilon^2 r \Omega$ and $v^z \sim \epsilon v^r$.

Taking into account these considerations, equation (\ref{finali}) gives (to lowest order in
$\alpha$ and $\epsilon$) the viscous damping rate
\begin{equation}
\frac{1}{\tau} \equiv \mbox{Im}(\Delta\si) = -\frac{\int(F_1+F_2+F_3+F_4)d\,^3x}{\int F_0d\,^3x}  \; . \label{thin}
\end{equation}
To obtain this expression, we employed (many) integrations by parts, with the boundary conditions that $\dVz$, $\eta$, and $\zeta$ vanish. The integrands are 
\begin{eqnarray}
F_0 & = & 2\om\rho\left(\frac{\rho}{\Gamma p}|\dVz|^2 + \kappa^2S^2\left|\frac{\partial\dVz}{\ppr}\right|^2\right) \; , \\
F_1 & = & \om[\om^2(2\eta+\xi)+\kappa^2\eta]\left|\frac{\partial}{\ppr}\left(S\frac{\partial\dVz}{\ppr}
\right)\right|^2 \; , \\
F_2 & = & \om^{-1}(2\eta+\xi)\left|\frac{\partial^2\dVz}{\ppz^2}\right|^2 \; , \\
F_3 & = & \om[\eta(\om^{-2} + S^2(\om^2+\kappa^2)) + 2S(\eta+\xi)] \left|\frac{\partial^2\dVz}{\ppr\ppz}\right|^2 \; , \\
F_4 & = & -\om S\left[(\mu-1)\frac{r\eta\rho}{\Gamma p}\left(\frac{\partial\Omega^2}{\ppr}\right) + \frac{\partial^2\xi}{\ppz^2}\right]
          \left|\frac{\partial\dVz}{\ppr}\right|^2 \; . 
\end{eqnarray}
We have introduced $S\equiv 1/(\om^2-\kappa^2)$ and $\xi\equiv\zeta-\case{2}{3}\eta$.  
The IL operator defined by equation (\ref{Ls}), which reduces to the form
\begin{equation}
\LL = \frac{\om^2\rho^2}{\Gamma p} 
+ \om^2\rho\frac{\partial}{\ppr}\left(S\frac{\partial}{\ppr}\right) 
+ \frac{\partial}{\ppz}\left(\rho\frac{\partial}{\ppz}\right) \; , \label{L0}
\end{equation}
has also been employed. 
It is interesting to note that the terms proportional to $v^r$ and $v^z$ do not appear in this limit, since their contribution is of higher order in $\epsilon$. This result justifies their neglect in previous analyses of thin disks.

Let us now consider in more detail the effect of viscosity on observationally relevant fundamental modes of relativistic accretion disks (near a black hole or neutron star). Since we are employing a Newtonian analysis, but use the properties of the relativistic eigenfunctions $\dVz$, our results are only approximate. The trapping of the g- and p-modes is produced by the fact that relativistic effects make the radial epicyclic frequency $\kappa$ less than the angular velocity of free-particle circular orbits. This reduction produces an inner edge of the disk where $\kappa(r)$ vanishes, with the maximum value of $\kappa(r)$ at a slightly larger radius. The trapping of the c-modes is related to the reduction of the vertical epicyclic frequency below the angular velocity, produced by the angular momentum of the central mass. 

\begin{itemize}

\item 
The g-modes are trapped where the comoving frequency $|\om| < \kappa$, near the maximum of $\kappa(r)$. They are {\it mathematically} similar to the internal gravity (buoyancy) modes of stars. For the fundamental ($m=0$) g-mode (\markcite{PSWL}Perez et al.~1997), $\pr\sim\epsilon^{1/2} \pz$, $\pz\sim 1/h_*$, and $\pf=0$ when applied to $\dVz$. Single terms in $F_0$ and $F_3$ dominate, giving  
\begin{equation}
\frac{1}{\tau} \cong -\frac{\int\eta(\si^2+\kappa^2)S^2|\pr\pz\dVz|^2 d\,^3x}
{2\int\rho\kappa^2 S^2|\pr\dVz|^2 d\,^3x} \sim - \alpha\Omega_* \; ,
\end{equation}
with $S\sim 1/(\epsilon\si^2)$.  
The sign of this result agrees with that obtained by Nowak \& Wagoner (\markcite{NW}1992), although the magnitude of their growth rate was greater.

\item 
The (inner) p-modes are trapped between the inner edge of the disk and the increasing portion of $\kappa(r)$, with $|\om| > \kappa$. They are similar to the pressure (acoustic) modes of stars. For the fundamental ($m=0$) p-mode (\markcite{NW1}Nowak \& Wagoner 1991), $\pr\sim\epsilon^{1/3} \pz$, $\pz\sim 1/h_*$, and $\pf=0$  when applied to $\dVz$. The numerator is dominated by the same term in $F_3$ as above, as well as $F_2$, giving  
\begin{equation}
\frac{1}{\tau} \cong -\frac{\int[(2\eta+\xi)\si^{-2}|\pz^2\dVz|^2 + \eta(\si^2+\kappa^2)S^2|\pr\pz\dVz|^2] d\,^3x}
{2\int[\rho c_s^{-2}|\dVz|^2 + \rho\kappa^2 S^2|\pr\dVz|^2] d\,^3x}
\sim -\frac{\alpha\Omega_*}{(\sigma/\Omega_*)^2 + \epsilon^{2/3}} \; ,
\end{equation} 
with $c_s^2=\Gamma p/\rho$ and $S\sim 1/\si^2$. The first (second) integral in both the numerator and the denominator dominates when $\epsilon$ is smaller (greater) than $(\sigma/\Omega_*)^3\ll 1$. The sign of this result agrees with that obtained in a local analysis by Kato et al. (\markcite{KFM}1998), although its form differs somewhat from theirs.

\item 
The c-modes are trapped between the inner edge of the disk and the radius where the Lense--Thirring frequency is the eigenfrequency. For the fundamental ($m=\pm 1$) c-mode (\markcite{SW}Silbergleit \& Wagoner 1999), $\pf = \pm 1$, $\pz\sim 1/h_*$ but $\pz^2 = 0$ when applied to $\dVz$. Thus this mode is vertically incompressible. One finds that 
\begin{equation}
\frac{1}{\tau} \cong -\frac{\int(F_3 + F_4) d\,^3x}{2\int\rho c_s^{-2}\om|\dVz|^2 d\,^3x}
\sim \pm\alpha\left(\frac{h_*}{\lambda_r}\right)^2\Omega_*  \; ,
\end{equation} 
where $S\sim\om^{-2}$. Note that this mode may either grow or be damped by viscosity.

\end{itemize}

We remind the reader that the real part of the corrections to all eigenfrequencies vanish.
Future generalizations of this work should include the effects of buoyancy (nonbarotropic equation of state) and effects of higher order in $\epsilon\equiv h_*/r_*$.

\acknowledgments

This research was supported in part by NASA grants NAG 5-3102 to R.V.W. and NAS 8-39225 to Gravity Probe B. We thank John Friedman, Lee Lindblom, and Alex Silbergleit for helpful suggestions.


\appendix
\section{APPENDIX: Determination of $\dR^a$ and $\dL^a_{\ b}$}

The components of $\nabla_{\! b} \sigma^{ab}$ in \occ \ are
\begin{mathletters}
\begin{eqnarray}
   \nabla_{\! b} \sigma^{rb} & = &
   2(\eta v^r_{,r})_{,r} + \oor [\eta (\oor v^r_{, \phi} + v^\phi_{,r} -
    \oor v^\phi)]_{,\phi} + [\eta(v^r_{,z} + v^z_{,r})]_{,z}  \nonumber \\
   & & \mbox{}+ 2\eta \oor (v^r_{,r} - \oor v^r - \oor v^\phi_{,\phi})
    + \{ (\zeta - \case{2}{3} \eta)
      [\oor(r v^r)_{,r} + \oor v^\phi_{,\phi} + v^z_{,z}] \}_{,r} \; ;  \\
  \nabla_{\! b} \sigma^{\phi b} & = &
  [\eta(v^\phi_{,r} - \oor v^\phi + \oor v^r_{,\phi})]_{,r}
  + 2 \eta \oor (v^\phi_{,r} - \oor {v^\phi} + \oor v^r_{,\phi})
  + 2 r^{-2} [\eta (v^r + v^\phi_{,\phi})]_{,\phi}  \nonumber \\
  & & \mbox{}+ [\eta(v^\phi_{,z} + \oor v^z_{,\phi})]_{,z}
  + \oor \{ (\zeta - \case{2}{3} \eta)
      [\oor(r v^r)_{,r} + \oor v^\phi_{,\phi} + v^z_{,z}] \}_{,\phi} \; ; \\
   \nabla_{\! b} \sigma^{zb} & = &
  [\eta(v^r_{,z} + v^z_{,r})]_{,r} + \eta \oor (v^r_{,z} + v^z_{,r})
  + 2 (\eta v^z_{,z})_{,z}  + \eta \oor (v^\phi_{,z}
  + \oor v^z_{,\phi})]_{,\phi}  \nonumber \\ 
  & & \mbox{}+ \{ (\zeta - \case{2}{3} \eta)
      [\oor(r v^r)_{,r} + \oor v^\phi_{,\phi} + v^z_{,z}] \}_{,z} \; . 
\end{eqnarray}
\end{mathletters}
These expressions and their perturbations allow us to compute the operator $\LLv$. There are two terms in the right-hand-side of equation (\ref{e2}) that contribute to $\alpha \, \dR$. 
We will consider each in order.
The first contribution comes from $-\rho^{-2} \drho \nabla_{\! b} \sigma^{a b}$.
The order $\alpha$ part of this contribution can be obtained from the above equations, noticing that $\pf = 0$ when acting on an equilibrium quantity and that the variables $v^r, v^z, \eta, \zeta, \deta$, and $\dzeta$ are all of order $\alpha$:
\begin{equation}
  \alpha \, \dR_A =
    -{1 \over \rho^2} \left[\matrix{
   0 \cr
    {(r \rho / p \Gamma)}
    (\eta \Omega_{,z})_{,z} + (\pr r + 2) {(\eta \rho \Omega_{,r} / p
     \Gamma)}
  \cr
   0 \cr}\right] \; . 
\end{equation}   
The second contribution comes from   
  $\rho^{-1} \nabla_{\! b}[\deta (\nabla^a v^b + \nabla^b v^a)]$:
\begin{equation}
  \alpha \, \dR_B =
    {1 \over \rho} \left[\matrix{
    {i m \eta \mu \Omega_{,r} / p \Gamma} \cr
     r \pz ({\eta \mu \Omega_{,z} / p \Gamma}) 
     + (\pr r + 2) ({\eta \mu \Omega_{,r} / p \Gamma})  \cr
   {i m \eta \mu \Omega_{,z} / p \Gamma} \cr}\right] \; . 
\end{equation}
The term $\rho^{-1}\nabla^a [(\dzeta - \case{2}{3} \deta) \nabla_{\! c} v^c]$   
is of higher order in $\alpha$ and can therefore be ignored. Thus 
\begin{equation}
  \alpha \, \dR =  \alpha \, \dR_A +  \alpha \, \dR_B \; . 
\end{equation}

For the determination of $\alpha\,\dL$, let us start with the contribution from the left-hand-side of 
equation (\ref{e2}). The components of the latter are
\begin{mathletters}
\begin{eqnarray}
\partial_t \dv^r + v^b \nabla_{\! b} \dv^r + \dv^b \nabla_{\! b} v^r 
& = & i \si \dv^r + i m \Omega \dv^r - 2 \Omega \dv^\phi
   + v^r \dv^r_{,r} + v^z \dv^r_{,z} \nonumber \\ 
& & \mbox{}+ v^r_{,r} \dv^r + v^r_{,z} \dv^z \; ,  \\
\partial_t \dv^\phi + v^b \nabla_{\! b} \dv^\phi + \dv^b \nabla_{\! b} v^\phi 
& = & i \si \dv^\phi + i m \Omega \dv^\phi +  \Omega \dv^r
   + v^\phi_{,r} \dv^r + v^\phi_{,z} \dv^z    \nonumber \\ 
& & \mbox{}+ v^r \dv^\phi_{,r} + v^z \dv^\phi_{,z} + \oor v^r \dv^\phi \; ,  \\
\partial_t \dv^z + v^b \nabla_{\! b} \dv^z + \dv^b \nabla_{\! b} v^z 
& = & i \si \dv^z + i m \Omega \dv^z + v^r \dv^z_{,r} 
   + v^z \dv^z_{,z} + v^z_{,r} \dv^r  \nonumber \\ 
& & \mbox{}+ v^z_{,z} \dv^z \; . 
\end{eqnarray}
\end{mathletters}
The first contribution to $\alpha\,\dL$ comes from the terms with $v^r$ or $v^z$:
\begin{equation}
   \alpha \, \dL_A = (v^r \pr + v^z \pz) \, {\bf{I}} \; + \;
    \left[ \matrix{ v^r_{,r} & 0 & v^r_{,z} \cr
    0 & {v^r / r} & 0 \cr
    v^z_{,r} & 0 & v^z_{,z} \cr} \right] \; , 
\end{equation}
where {\bf{I}} is the identity matrix. Let us turn now to the contributions to $\alpha\,\dL$ coming from the right-hand-side of equation (\ref{e2}). 
The first of such contributions comes from the term ${\rho}^{-1}\nabla_{\! b}[\eta(\nabla^a \dv^b + \nabla^b\dv^a)]$:
\begin{displaymath}
\alpha \, \dL_B = {1 \over \rho} 
   \left[
     \matrix{
      2 \pr \eta \pr + 2 \eta \pr \oor     
        - m^2 r^{-2} \eta + \pz \eta \pz
        &
     {i m \eta} \oor (\pr - {3 \oor})      
        \cr
     (\pr + {4 \oor}) {i m \eta \oor}
        &
       (\pr r + 2) \eta \pr \oor - {2 m^2 r^{-2} \eta}
         + \pz \eta \pz
        \cr
      (\pr + \oor) \eta \pz
        &
      {i m \eta \oor} \pz
        \cr
    } \right.   
  \end{displaymath}
  \begin{equation} \left.
     \matrix{
      \pz \eta \pr
         \cr
      {i m \oor} \pz \eta
        \cr
      (\pr + \oor) \eta \pr + 2 \pz \eta \pz - {m^2 r^{-2} \eta}
        \cr } \right] \; . 
\end{equation}
The other contribution comes from the term $\rho^{-1}\nabla^a [(\zeta - \case{2}{3}\eta) \nabla_{\! c} \dv^c]$:
\begin{equation}   
   \alpha \, \dL_C = {1 \over \rho}\left[\matrix{
   \pr {\xi} \oor \pr r & i m \pr {\xi} \oor  & \pr \zme \pz  \cr
   {i m \xi} r^{-2} \pr r& - {m^2} r^{-2} \zme
       & {i m \xi} \oor \pz \cr
   \pz {\xi} \oor \pr r  & {i m} \oor \pz \zme
       & \pz \zme \pz   \cr}\right] \; , 
\end{equation}
where $\xi \equiv \zeta - \case{2}{3} \eta$. Thus
\begin{equation}
   \alpha \, \dL =  \alpha \, \dL_A +  \alpha \, \dL_B +  \alpha \, \dL_C \; . 
\end{equation}


\end{document}